# Observation of Weyl and Dirac fermions at smooth topological Volkov-Pankratov heterojunctions

J. Bermejo-Ortiz[1,†], G. Krizman[2,*,†], R. Jakiela[3], Z. Khosravizadeh[3], M. Hajlaoui[2], G. Bauer[2], G. Springholz[2], L.-A. de Vaulchier[1], Y. Guldner[1]

[1] *Laboratoire de Physique de l'Ecole normale supérieure, ENS, Université PSL, CNRS, Sorbonne Université, 24 rue Lhomond 75005 Paris, France*

[2] *Institut für Halbleiter und Festkörperphysik, Johannes Kepler Universität, Altenbergerstrasse 69, 4040 Linz, Austria*

[3] *Institute of Physics, Polish Academy of Sciences, Warsaw, Poland*

*Corresponding author: gauthier.krizman@jku.at
†These authors contributed equally to this work

**Weyl and Dirac relativistic fermions are ubiquitous in topological matter. Their relativistic character enables high energy physics phenomena like the chiral anomaly to occur in solid state, which allows to experimentally probe and explore fundamental relativistic theories. Here we show that on smooth interfaces between a trivial and a topological material, massless Weyl and massive Dirac fermions intrinsically coexist. The emergence of the latter, known as Volkov-Pankratov states, is directly revealed by magneto-optical spectroscopy, evidencing that their energy spectrum is perfectly controlled by the smoothness of topological interface. Simultaneously, we reveal the optical absorption of the zero-energy chiral Weyl state, whose wavefunction is drastically transformed when the topological interface is smooth. Artificial engineering of the topology profile thus provides a novel textbook system to explore the rich relativistic energy spectra in condensed matter heterostructures.**

## I. INTRODUCTION

Weyl and Dirac electrons have attracted tremendous interest as they mimic relativistic high-energy particles. The properties of these electrons are governed by physical laws that resemble relativistic equations [1–3] and as a result, condensed matter and relativistic physics have emerged as tightly related fields. A striking example is the discovery of topological insulators (TI), which are insulating in the bulk but exhibit a metallic chiral state (CS) at their boundaries [4–6]. These massless states obey the Weyl equation [1,7,8] and display a relativistic energy-momentum relation. As such, they are a hallmark of relativistic physics inherent to topological matter, and with the advent of topological systems, they have been coined "topological interface states" (TIS). They display a distinct chirality with the electron spin locked to its momentum protected by time reversal or crystalline symmetry. This led to the discovery of novel phenomena such as the quantum spin Hall (QSH) [9,10] and quantum anomalous Hall effects [11,12] and opened up the doors for novel physics and device applications.

The fundamental origin of the TIS is the change of band topology an electron experiences at the boundary between a normal insulator (NI) and a TI. The generic system to showcase the TIS existence is thus, a topological heterojunction (THJ), that is, a single NI/TI interface – gradual or abrupt – where a crossover of topology from a material with positive to one with negative, i.e., inverted band gap occurs. Dictated by the fundamental bulk-boundary correspondence principle that two topologically distinct regions cannot be connected or transformed from one to another without a band crossing at the interface, this intrinsically leads to the formation of a gapless interface state. In most general terms, the topological crossover at a THJ can be widely extended in space and can consist of an arbitrary varying band gap profile, as long as a distinct change of a certain topology invariant takes place. For this very reason, topological CS emerge *independently of how* the band inversion evolves across the interface region. Although this fundamental prediction has been disclosed by Volkov and Pankratov almost 40 years ago [7,8] it has not been actually tested/proven by experiments. This is, on the one hand, due to the considerable technological challenge of fabricating *gradual THJs* with artificially varying band gap profiles produced on demand by band gap engineering; on the other hand, unlike the topological states formed at TI surfaces, the chiral interface state of a gradual THJ resides in a region enclosed by a NI and a TI material and as such, is not accessible by the usual surface sensitive probes like angle resolved photoemission spectroscopy (ARPES) or scanning tunneling microscopy.

In this work, we employ band structure engineering and molecular beam epitaxy to create artificial THJs with perfectly controlled smooth band gap profiles, crossing zero along the junction direction. Probing the relativistic energy spectrum by magneto-optical spectroscopy, we show that the topological CS forms a perfectly robust zero energy mode that is completely independent of the smoothness of the THJ transition. In particular, the chiral zero energy mode persists for THJ as wide as 200 nm, which impressively validates the fundamental prediction of topologic band theory [4,7,13]. Our artificial THJ are created using the lead tin chalcogenide $Pb_{1-x}Sn_xSe$ system in which the fundamental band gap $2\Delta$ smoothly evolves from a positive to a negative value as the Sn concentration increases beyond a certain critical value $x_c$ [14–16]. Accordingly, below $x_c$ the material is a normal insulator whereas above a topological crystalline insulator (TCI) with inverted band gap forms as signified by the appearance of a gapless topological surface state as shown by Fig. 1(a-c). By digital control of the alloy composition, we create smooth Volkov-Pankratov THJs (VP-THJ) with arbitrary junction widths to engineer on demand the transition $2\Delta(z)$ from the NI to the TCI state.

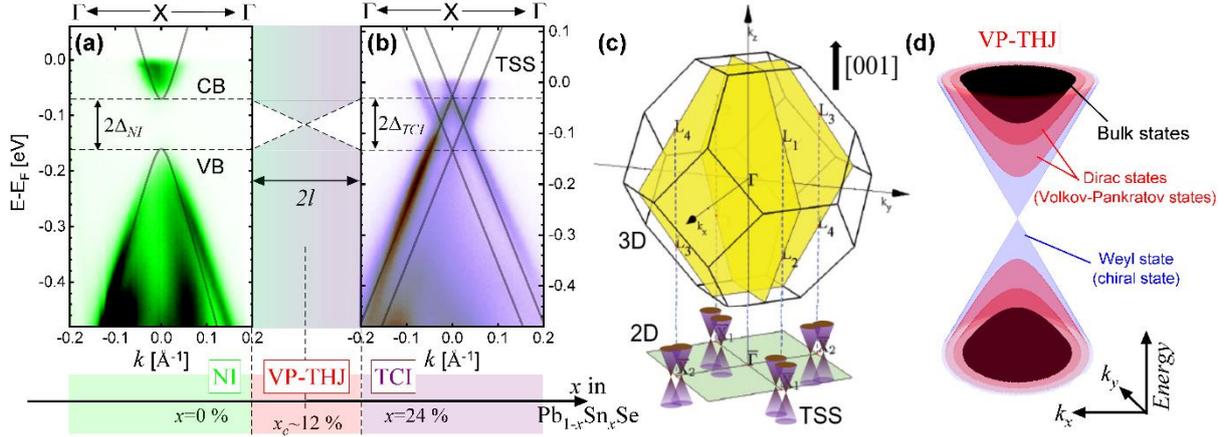

**FIG. 1. Smooth topological heterojunction with the $Pb_{1-x}Sn_xSe$ system.** Energy-momentum spectra of the NI PbSe **(a)** and the TCI $Pb_{0.76}Sn_{0.24}Se$ **(b)** measured by ARPES at $T = 10$ K, at the surface of epitaxial layers grown on (001) KCl substrates. Note that the samples were intentionally highly *n*-doped to shift the Fermi energy deep into the conduction band. The analysis yields a band gap of $2\Delta_{NI}= +90$ meV and $2\Delta_{TCI}= -95$ meV for (a) and (b), respectively (dashed lines), evidencing the band inversion. The solid lines in (a) represent the $E(k)$ dispersions of the PbSe bulk bands and in (b) the Dirac cones of the TSS of $Pb_{0.76}Sn_{0.24}Se$ as derived by $k.p$ theory. By grading of composition $x$ the band gap can be smoothly tuned from positive to negative to form VP-THJs with well-defined widths $2l$ as illustrated between the panels (a) and (b). **(c)** Bulk and (001) surface Brillouin zones of $Pb_{1-x}Sn_xSe$. In the bulk, the band inversions occur at the $L$-points, which are projected pairwise to the $\overline{X}$-points (red dots) on the (001) surface. The coupling within each pair splits the TSS into a double Dirac cone due to the Lifshitz transition as seen in (b). **(d)** Typical 2D energy spectrum of a sufficiently smooth VP-THJ, where massless Weyl (blue) and massive Dirac (red) fermions are located energetically within the bulk gap.

Probing the confined 2D states of smooth THJ reveals the emergences of a new relativistic energy spectrum that features massive Dirac fermions coexisting with the massless Weyl (the CS) fermion at a single interface, as illustrated by Fig. 1(d). While any THJ always supports *one* massless CS with definite chirality, here we show that the additional massive states emerge only when the THJs become sufficiently wide. Following the prediction of Volkov-Pankratov theory [7,8] these states are called Volkov-Pankratov states (VPS) [7,8,17–19]. Like the CS, the VPS are also chiral in nature and intrinsically relativistic but display a *massive* Dirac energy-momentum dispersion relation. The VPS display a natural electron-hole symmetry and although they are gapped, they lie within the band gaps of the confining TCI and NI regions. Moreover, their wave functions deeply extend into topological trivial and non-trivial regions and as such, they fundamentally differ from classic confined 2D states in normal semiconductor heterojunctions, which encode, by a certain quantization, only the NI bulk band dispersions. By our experiments, we determine the critical value of the THJ width for the appearance of the VPS states and reveal that thereafter the number of VPS increases, meaning that, contrary to the CS, multiple VPS can be supported in a VP-THJ. We particularly highlight the unique scaling properties of the VPS in dependence of the junction width shown to be in perfect quantitative agreement with the VP theory [7,17]. Our VP-THJ thus, represent a textbook system in which relativistic Weyl and Dirac fermions can be engineered in coexistence, which opens up new avenues for novel physics and device applications.

## II. DESIGN AND FABRICATION OF VOLKOV PANKRATOV THJs

The formation of Weyl and Dirac fermions in gradual topological heterojunctions demands a material system that allows a precise engineering of the band topology profile across the junction region. In this work, we employ the lead tin chalcogenide system, specifically, THJ based on $Pb_{1-x}Sn_xSe$ as originally envisioned in the Volkov-Pankratov work [7]. These belong to the IV-VI topological crystalline insulator (TCI) class of materials [18–22] that features a smooth band inversion from NI to TCI when the Sn content increases [14] and for this reason is perfectly suited for topological band gap engineering. The IV-VI semiconductors have been extensively used for fabrication of quantum confined structures such as quantum wells [23–26] and quantum dots [27,28] used for a wide range of optoelectronic devices such as mid-infrared laser diodes [29,30], vertical cavity surface emitting lasers [31] as well as focal plane arrays [25]. Moreover, carrier mobilities as high as 100.000 $cm^2$/Vsec can be reached in $Pb_{1-x}Sn_xSe$ [32] and the material can be made $n$- or $p$-type by impurity doping. Furthermore, IV-VI TCIs can support various topological phenomena such as Weyl nodes with ferroelectric distortions [33,34] or magnetic doping [35], 1D step-edge states [36,37], as well as the QSH effect [38,39].

The $Pb_{1-x}Sn_xSe$ band structure is characterized by nearly mirror symmetric conduction and valence bands with the band gaps located at the $L$-points of the Brillouin zone (BZ) [40,41] (cf. Fig. 1(c)). In the vicinity of these points the conduction and valence bands are well described by $\boldsymbol{k}.\boldsymbol{p}$ theory [40–42], and in good approximation the fundamental band gap $2\Delta$ linearly changes with Sn concentration $x_{Sn}$ according to

$$2\Delta = 2\Delta_0(T, \varepsilon) - \alpha \cdot x_{Sn} \qquad (1)$$

where $\alpha \sim 800\ meV$ [14,16,43] and $\Delta_0$ is the band gap of PbSe at a given temperature $T$ and strain state $\varepsilon$ that can be controlled by the epitaxial growth conditions. Accordingly, $\Delta$ changes from positive to negative at a critical Sn concentration $x_c = 2\Delta_0/\alpha$ at which the topological phase transition from the NI to the TCI state takes place. This is illustrated by Fig. 1(a,b), where energy-momentum maps of $n$-doped PbSe and $Pb_{0.76}Sn_{0.24}Se$ on KCl (001) substrates recorded by ARPES around the $\bar{X}$-point of the (001) surface BZ are depicted. Evidently, the band gap of NI PbSe is open, while that of the topological surface of $Pb_{0.76}Sn_{0.24}Se$ is closed, which directly evidences the non-trivial band topology and band inversion in this case.

For PbSe the conduction and valence bands are in perfect agreement with $\boldsymbol{k}.\boldsymbol{p}$ theory [26,38,40,41] represented by the solid lines in Fig. 1(a). Note that other energy bands lie about 2 eV above and below the neutral point [15]. Therefore, the 2-band $\boldsymbol{k}.\boldsymbol{p}$ Hamiltonian is almost unperturbed, which renders $Pb_{1-x}Sn_xSe$ a nearly perfect Dirac material similar to graphene. For the TCI $Pb_{0.76}Sn_{0.24}Se$ two Dirac cones appear at the surface [16,44] due to the pairwise projections of the symmetry inversions at the $L$-points of the bulk 3D Brillouin (BZ) to the $\bar{X}$-points of the (001) surface BZ as shown in Fig. 1(c). This leads to a splitting of the Dirac cones by the Lifshitz transition [16,45,46]. The inverted bulk band gap $Pb_{0.76}Sn_{0.24}Se$ is given by the separation of the Lifshitz points at $\bar{X}$ ($k_\parallel = 0$) [16,45,46] where the two Dirac cones cross each other, as is indicated by the solid lines in Fig. 1(b) [16,45,46]. This yields the band gaps of PbSe and $Pb_{0.76}Sn_{0.24}Se$ as $2\Delta_{NI} = +90 \pm 10$ meV and $2\Delta_{TCI} = -95 \pm 10$ meV epilayers at 10 K. Note that due to the epitaxial strain induced by the KCl substrate, these values differ from the bulk values [14,47] - the details of this will be reported elsewhere. Interpolation between these band gap values yields a critical Sn concentration $x_c \sim 12\%$ (see Eq. (1)) at which the band inversion takes place in our epilayers. By gradually increasing the Sn content from 0 to 0.24 during MBE growth, we can thus, produce perfectly symmetric Volkov-Pankratov THJs with the band inversion point located exactly in the middle of junction as shown schematically in Fig. 1(a,b).

To create the graded PbSe/Pb$_{1-x}$Sn$_x$Se THJ with well controlled compositional profiles in the junction region, we employ molecular beam epitaxy of digitally controlled ultra-short period PbSe/Pb$_{0.76}$Sn$_{x0.24}$ superlattices (SL) in which the thickness ratio and thus, the SL composition was incrementally changed over the junction region. This allows to produce textbook-like THJs on demand with artificially engineered band profiles and topological transitions spanning between regions of large inverted/non-inverted band gaps ($\sim \pm 95$ meV). The growth starts with a 200 nm-thick NI PbSe buffer layer on a (001) KCl substrate, followed by the digitally graded Pb$_{1-x}$Sn$_x$Se junction region with linearly increasing Sn content from $x_{Sn} = 0$ to 24 % in fine steps, as is ensured by the large number of up to 120 ultra-thin layers in the junction region as detailed in Appendix A. All THJ s are terminated by a 200nm Pb$_{0.76}$Sn$_{0.24}$Se TCI layer with constant composition. A whole series of THJs with varying thickness $2l$ of the graded junction region was prepared, starting from $2l = 0$ (abrupt THJ) to $2l = 200$ nm, which is the widest THJ investigated in this work. For clarity, the samples are labelled as "THJ-$2l$" where $2l$ corresponds to the thickness (in nm) of the smoothly graded interface layer.

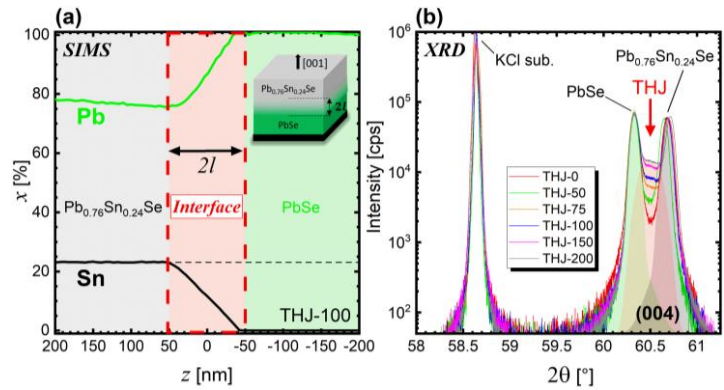

**FIG. 2. Structural characterization of the VP-THJ. (a)** SIMS depth profile of the composition for the PbSe/Pb$_{0.76}$Sn$_{0.24}$Se THJ-100 measured along the [001] growth direction, evidencing the linear change of the Pb and Sn concentration over the graded interface region. **(b)** X-ray diffraction spectra (using CuKα1 radiation) around the (004) Bragg peak of the six investigated THJs with $2l$ varying from 0 up to 200 nm. The peaks from the KCl substrate, PbSe, and Pb$_{0.76}$Sn$_{0.24}$Se are indicated. The diffraction curves are normalized to the PbSe peak.

The compositional profiles of the THJs were determined by secondary ion-mass spectroscopy (SIMS) [48]. Figure 2(a) shows the measured depth profiles of the Pb and Sn atom concentrations for the THJ with 100 nm wide junction region (THJ-100). As one can see, indeed the grading is essentially linear and the total thickness of the junction region perfectly agrees with the targeted value. The sample structure was further analyzed by high resolution X-ray diffraction as shown in Fig. 2(b). For the graded THJ, the diffraction spectra feature two main peaks stemming from the 200 nm thick bottom PbSe and top Pb$_{0.76}$Sn$_{0.24}$Se layers with constant composition. The graded junction region shows up as a wide and almost constant intensity distribution in between these peaks (red shading in Fig. 2b) that is due to the continuously changing out-of-plane lattice parameter in the graded region that decreases from the value $a = 6.130$ Å of PbSe to 6.100 Å for Pb$_{0.76}$Sn$_{0.24}$Se. Evidently, its intensity rises proportional to the thickness of the graded junction region (cf. Fig. 2(b)) and is thus a direct measure of the THJ width. Using the Vegard's law [32] to determine the composition of the boundary layers of the THJ, we again find an excellent agreement with the targeted values, which proves the excellent control and reproducibility of our growth.

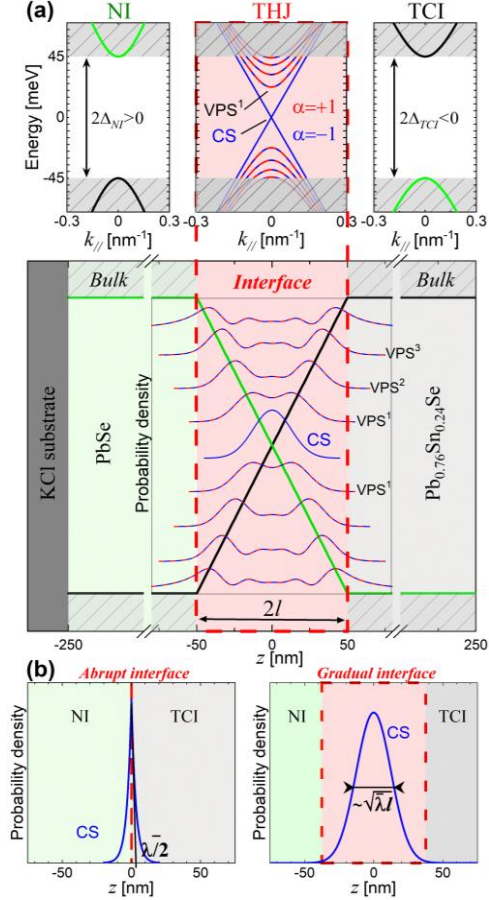

**FIG. 3. Weyl and Volkov-Pankratov states formed in a smooth THJ.** The THJ is illustrated for the case of a NI/TCI junction, consisting of NI PbSe on the one side and band inverted TCI Pb$_{0.76}$Sn$_{0.24}$Se on the other side that are connected by a 100 nm thick linearly graded topological junction region where the Sn content increases from 0 to 24 % (THJ-100 sample). **(a)** Top: Band diagrams (in-plane dispersions in the reciprocal space) for the three regions NI, THJ region, and TCI from left to right. The dispersions of the interface states are derived within a chiral basis $\alpha = \pm 1$ in red or blue. Middle: Real space band edge profile along the $z//[001]$ direction illustrating the smooth inversion of the conduction and valence bands (green and black lines respectively). Superimposed are the calculated envelope wavefunctions of the massless CS and massive VP states in the chiral basis. **(b)** Comparison of the envelope wavefunctions of the CS for the case of an abrupt (left) and a gradual (right) interface. For the abrupt interface, the black line denotes the tangent at $z = 0$, which defines the characteristic length $\bar{\lambda}/2$. For the gradual interface, the CS wavefunction has a Gaussian shape with a FWHM given exactly by $2\sqrt{2\ln 2}\sqrt{\bar{\lambda}l}$.

Using magneto-optical spectroscopy we accurately determined the band gaps of the NI and TCI ends of the THJ (see Appendix B), giving $2\Delta_{NI} = 2\Delta(z < -l) = +95$ meV for the bottom PbSe and $2\Delta_{TCI} = 2\Delta(z > +l) = -95$ meV for the top Pb$_{0.76}$Sn$_{0.24}$Se end of the VP-THJs, in perfect agreement with our ARPES measurement (Fig. 1). The $2l$ width of the graded THJ region defines the slope of the band gap profile $\partial 2\Delta/\partial z = 2(\Delta_{NI} - \Delta_{TCI})/l$, giving a value of 1 - 10 meV/nm for the THJs with $2l = 20 - 200$ nm, respectively. The resulting THJ band gap profile is illustrated by Fig. 2a for 100 nm junction width, indicating that due to the symmetric structure the topological transition $2\Delta(z) = 0$ is in the middle of the structure. The designed gap function $\Delta(z)$ was used as input parameter for calculation of the relativistic eigenstates of the THJ using $\boldsymbol{k}.\boldsymbol{p}$ theory described in Appendix C. The results are shown in the top panel of Fig. 3(a) for the THJ-100 by the $E(k)$ dispersion curves of the bulk and TIS for the three NI, THJ and TCI regions.

As shown in the middle panel of Fig. 3(a), for a smooth THJ, a series of confined interface states with clearly defined chirality $\alpha = \pm 1$ are predicted. These states consist of the gapless chiral Weyl state (CS) with zero mass and perfectly linear $E(k)$ dispersion, as well as the additional VPS that are gapped and display a finite mass [1,7,8,14–16]. According to our calculations, for a linearly graded THJ they emerge exclusively under the conditions that the junction width $2l$ exceeds a critical value that is proportional to the characteristic length $\bar{\lambda} \sim \hbar\bar{v}_z/\bar{\Delta}$ defined by the averaged Dirac velocities $\bar{v}_z$ and band gaps $\bar{\Delta} > 0$ of the two endings of the THJ region. In fact, the number of VPS is given by $\lfloor l/2\bar{\lambda} \rfloor$ as shown in detail below. A particularly interesting point is that the THJ grading strongly affects the wave function envelope of the CS as is illustrated by Fig. 3(b). Although its maximum is always at the band crossing point, for an abrupt interface, the CS wave function is cusped and rapidly decays exponentially on each side of the junction with a decay length $\bar{\lambda}/2$. On the contrary, for a gradual THJ, the CS wavefunction assumes the form of a Gaussian (see Fig. 3(b)) with a full width at half maximum (FWHM) proportional to $\sqrt{\bar{\lambda}l}$. Thus, the CS extends over a much wider region of space in a VP-THJ [17].

### III. MAGNETO-OPTICAL SPECTROSCOPY OF THE WEYL AND DIRAC STATES

Having set the stage to study the relativistic spectrum of VP-THJ, we now turn to magneto-optical spectroscopy to detect the CS and VPS properties for different interface thicknesses. Figure 4(a) shows the magneto-optical spectra of the samples THJ-0, THJ-50, THJ-75, THJ-100, THJ-150 and THJ-200 measured at $T = 4.2$ K and magnetic field $B = 15$ T. In each case, we clearly observe distinct absorptions (blue and red arrows in Fig. 4(a)) at energies below the bulk band gaps $2|\Delta_{NI,TCI}| \approx 100$ meV. Strikingly, the lowest energy absorption marked in blue shows up at the same position for all THJs independently of their interface width, whereas the higher energy absorptions, marked in red, shift to lower energies when the heterojunction width $2l$ increases. This is a first clear indication that these absorptions arise from the VPS.

To quantitively analyze the data, let us model the gradual interface using a half-gap function $\Delta(z)$ that smoothly changes along $z//[001]$. Using the theory developed by Volkov and Pankratov [7,8], and by Lu and Goerbig [17], we express the Dirac Hamiltonian describing Pb$_{1-x}$Sn$_x$Se alloys in the Weyl basis. Then, the Hamiltonian of the gradual interface is derived and gives the following chiral-polarized equation at $k_\parallel = 0$:

$$E^2 \chi_\alpha = [-\hbar^2 v_z^2 \partial_z^2 + \Delta^2(z) + \alpha \hbar v_z \partial_z \Delta(z)]\chi_\alpha \quad (2)$$

where $\chi_\alpha$ is a component of the spinor, having a well-defined chirality $\alpha = \pm 1$. Note that Eq. (2) focuses on the squared energy. Besides the chirality, the solutions of Eq. (2) evidently depend on the potential shape $\Delta(z)$ and thus, this equation can model different relativistic spectra in condensed matter. The hereby investigated THJs display only the simplest case of a linear profile $\Delta(z)$, as shown in Fig. 2(a) and 3(a). The square energy spectrum for such linearly graded THJ, at $B = 0$ or $k_\parallel = 0$, turns out to be simply the one of a 1D quantum harmonic oscillator [17], and thus the energy spectrum can be written as:

$$E_{n,\alpha}(B = 0) = \pm\bar{\Delta}\sqrt{\frac{2\bar{\lambda}}{l}\left(n + \frac{1+\alpha}{2}\right)} \quad (3)$$

where the quantum number $n \geq 0$ is an integer and $\alpha = \pm 1$ is the chirality of the state. $\bar{\Delta} > 0$ represent the average half band gap $\bar{\Delta} = (\Delta_{NI} - \Delta_{TI})/2$ of the THJ, which in our case is ~45 meV for all samples (see Table I in Appendix B). Likewise, $\bar{\lambda}$ is the averaged characteristic length associated to the two ends of the THJs: $\bar{\lambda} = \hbar\bar{v}_z/\bar{\Delta}$. Using the average Dirac velocity $\bar{v}_z = 4.20\times10^5$ m/s along the growth direction as obtained in our previous work [14], one finds $\bar{\lambda}\sim 6$ nm. We highlight that the 1D harmonic oscillator squared-energy spectrum of the THJ (Eq. (3)) directly emerges from the linear $\Delta(z)$ potential that can be therefore seen as a pseudo-magnetic field with associated pseudo-magnetic length written as $\sqrt{\bar{\lambda}l}$ [49]. Consequently, the eigenfunctions $\chi_\alpha$ are Hermite polynomials, which are shown as the squared wavefunctions plotted in Fig. 3(a). A resolution using the envelope function approximation is presented in Appendix C and leads to the same results. The envelope function theory, however, stands as a more general approach that can be easily applied to any arbitrary potential $\Delta(z)$.

For the ground interface state with $n = 0$ in Eq. (3), we retrieve the massless topological CS with $E_{0,-} = 0$ that displays a precisely defined chirality ($\alpha = -1$). The other solutions for $n > 0$ describe the additional twofold degenerate VPS existing in a gradual THJ, whose energies lie within the band gap of the boundary layers. Note that the lowest energy VPS ($n = 1$ for chirality $\alpha = -1$, and $n = 0$ for $\alpha = +1$) are bound only if $l > 2\bar{\lambda}$, i.e., only if the interface is sufficiently gradual, whereas the CS persists even for an abrupt interface, as it intrinsically results from fundamental band symmetry inversion across the interface. Note that additional details such as a possible gap asymmetry or a different band alignment and/or doping of the NI and TCI sides of the THJ are described in Appendix D and in Ref. [50].

When a magnetic field ***B*** is applied along the [001] direction, the Weyl and Dirac dispersions become fully quantized forming relativistic Landau levels as sketched in Fig. 4(c) with energies given by [49]:

$$E_{n,\alpha,N,\pm}(B) = \pm\sqrt{E_{n,\alpha}^2(B=0) + 2e\hbar\bar{v}_\parallel^2 BN} \qquad (4)$$

Here, $N = 0, 1, \ldots$ is the Landau level index and $\bar{v}_\parallel$ the averaged in-plane velocity of the trivial NI and topological TCI layers, giving $\bar{v}_\parallel \approx (v_{\parallel,NI} + v_{\parallel,TI})/2 \sim 4.90\times10^5$ m/s using the values derived in Table I in Appendix B. Note that Eq. (4) describes pure relativistic Landau levels. The four valleys present in these compounds, located at the $\bar{X}$ points of the 2D BZ (see Fig. 1(c)), are all equivalent with respect to the [001] direction. They all verify Eq. (4) and thus, contribute to the same magneto-optical absorptions.

Let us first focus on the lowest energy magneto-optical absorptions of the THJs marked in blue in Fig. 4(a). Their energetic positions are, within the experimental accuracy (1-2 meV), independent of the interface thickness. Measuring their magnetic field dependences yields the blue dots in the Landau fan charts that are presented in Fig. 4(b) for all five gradual THJs. Using Eq. (4), the blue data points are fitted by the ground interband transitions of the CS indicated by the blue arrow in Fig. 4(c), involving the Weyl state ($n = 0$ and $\alpha = -1$), giving the blue solid lines in Fig. 4(b). Evidently, these perfectly fit the experiments, giving $\bar{v}_\parallel = 5.10 \pm 0.15$ m/s independently of the THJ interface thickness. This value agrees very well to the average $\bar{v}_\parallel$ calculated from the spectra of the bulk parts of the THJ.

Interestingly, the CS is not observed in the sample THJ-0 with the sharp interface. This is attributed to the Lifshitz transition when the interface is abrupt, as seen in ARPES at a TCI/vacuum interface (see Fig. 1(b)). Under this condition, the valley interaction between the two cones is responsible for the emergence of off-diagonal terms in the Hamiltonian given in Ref. [46] that mix the spin and orbital characters of the two cones. The overall selection rules are therefore relaxed, and the absorption is strongly weakened as compared to the single cone picture. This means that for the abrupt THJ-0 the signature of the CS falls below the noise level of our experiments. In the case of smooth interfaces, scattering between two valleys that project at the same in-plane momentum (see Fig. 1(c)) requires a large out-of-plane momentum transfer and thus, vanishes quickly when $l$ increases [46]. For the graded THJs, the absorption coefficient of the CS is therefore proportional to the fine structure constant, similar as in graphene [51], and as a result, a stronger and sharper absorption is obtained.

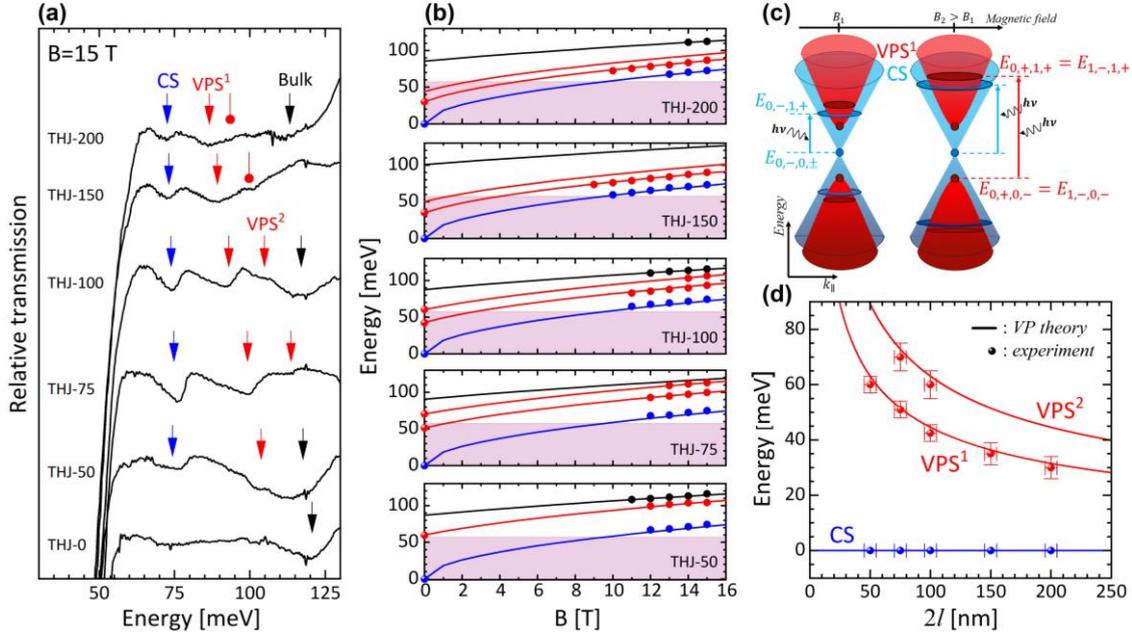

**FIG. 4. Magneto-optical evidence of the VP-THJ energy spectra. (a)** Normalized transmission spectra at $B = 15$ T and $T = 4.2$ K for the six THJs with different widths of the topological interfaces. The low energy absorptions marked by blue, red and black arrows denote the optical transitions involving the CS, VPS, and bulk respectively. **(b)** Landau level fan charts of the absorptions (dots) plotted together with the fit of the ground transitions using Eq. (4). Blue is for the CS; VPS[1] and VPS[2] are in red; and the ground interband transitions of bulk Pb$_{0.76}$Sn$_{0.24}$Se is plotted in black. The shaded purple region indicates the experimentally inaccessible energy range. **(c)** Quantized band structure of CS and VPS[1] in a perpendicular magnetic field, shown for two arbitrary magnetic fields $B_1$ and $B_2$. The Weyl and Dirac cones are quantized in terms of Landau levels with energies $E_{n,\alpha,N,\pm}(B)$. The blue and red arrows indicate the magneto-optical CS and VPS[1] ground transitions seen in panel (a). The extrapolation of the transitions to zero magnetic field yields the ground state energies $2E_{n,\alpha}(B = 0)$ of the CS and VPS. These energies are plotted versus $2l$ in **(d)** The solid lines represent $2E_{n,\alpha}(B = 0)$ obtained by Eq. (3), without any fit parameters.

In contrast to the observed chiral Weyl state transitions, the absorptions at higher energies, indicated by the red arrows in Fig. 4(a), show a very different behavior as their position strongly depends on the junction thickness. This is because these transitions arise from the VPS, whose energies vary in dependence of the THJ width according to Eq. (3). Indeed, in the Landau fan charts, the experimental VPS[1] and VPS[2] absorptions, represented by the red dots in Fig. 4(b), are again in perfect agreement with the transitions calculated using Eq. (4) that are represented by the red solid lines in Fig. 4(b) and allow for the accurate determination of $2E_{n,\alpha}(B = 0)$. The corresponding ground interband transition of VPS[1] is schematically indicated by the red arrow in Fig. 4(c). We obtain an excellent fit between theory and experiments, using $\bar{v}_\parallel = 4.90 \pm 0.10$ m/s which is naturally equal to the averaged value across the heterojunction. Remarkably, even higher order VPS ($n = 2$) can be resolved in THJ-75 and THJ-100, whereas it is absent for the thinner THJ-50. This result agrees with the theory because the $2l = 50$ nm junction is simply not sufficiently thick to support any higher order VPS, the number of VPS being $\lfloor l/2\bar{\lambda} \rfloor$. On the other hand, because for the smooth THJ-150 and THJ-200 the VPS transitions become very close to each other, they are barely resolvable by our experiments (see point-arrows in Fig. 4(a)).

As shown by Fig. 4(b), the extrapolation of the transitions to zero magnetic field yields the ground state energies $2E_{n,\alpha}(B=0)$ of the relativistic states, as defined by Eq. (3). The resulting values, which correspond to the gap between the hole and electron-like CS and VPS$^n$, are plotted in Fig. 4(d) as a function of the THJ interface width, evidencing the anticipated decrease of the VPS energy with increasing junction width as well as the constant zero-gap energy for the CS. The theoretical predictions depicted by Eq. (3) are represented by the solid lines in Fig. 4(d), giving a striking evidence for the perfect agreement between theory and experiments for the whole series of THJs without adjustable parameters. Our experiments impressively demonstrate the $E_{n,\alpha} \sim \sqrt{n/l}$ dependence of the VPS, which validates quantitatively the Volkov-Pankratov and Lu and Goerbig theories [7,17] even with a remarkable accuracy. Moreover, we experimentally demonstrate a fundamental property of topology, namely, the immutable presence of the CS as long as a band crossing occurs. Last but not least, it is noted that relativistic CS and VPS transitions are not observed when the magnetic field is in the layer plane of the THJ (see Fig. 10 in Appendix E), which confirms the two-dimensional character of the interface CS and VPS [49].

## IV. DISCUSSION

As described above, VPS only appear in smooth THJs in which the band gap profile $\Delta(z)$ varies slowly over a wide length scale $l > 2\bar{\lambda}$. Thus, they do not exist when the THJ is abrupt. However, other types of massive 2D states can appear in abrupt THJs, originating from different mechanisms. For instance, they can be induced when an abrupt THJ is subjected to a large electrostatic band bending potential superimposed over an abrupt gap change at the interface as has been shown, e.g., by ARPES for TI surfaces such as $Bi_2Se_3$ where by surface doping induced by molecular gas adsorption or alkaline element deposition [52–54] a triangular potential well is formed by the resulting band bending at the surface, leading to the appearance electrostatically confined states (ECS).

Two-dimensional massive states can also naturally emerge at abrupt THJ made from Kane materials, like in HgTe/HgCdTe heterointerfaces [55–57]. In this case, the presence of the heavy hole band is responsible for an intricate electronic density at the surface [58–60]. The TIS, whose presence is driven by the $\Gamma_6$-$\Gamma_8^l$ inverted band gap, shows a dispersion that is strongly hybridized with heavy hole states ($\Gamma_8^h$) and thus, does not display a Weyl-like dispersion like the CS in VP-THJs observed in this work [56,59]. Moreover, the TIS in Kane materials resides in the $\Gamma_6$-$\Gamma_8^l$ inverted gap, and thus, is buried far below the $\Gamma_8^h$ band. An electrostatic potential is then needed to shift the CS into the experimentally reachable $\Gamma_8^h$-$\Gamma_8^l$ gap (slightly opened by a tensile strain) [55,56,61], a mechanism that is shown in Fig. 9 of Appendix D. The additional massive states that such an abrupt interface can host are known as the Dyakonov-Khaetskii's states (DKS) [58]. They are solutions to the Luttinger Hamiltonian, which depicts the $\Gamma_8^h$-$\Gamma_8^l$ subspace only, thus, their dispersions are parabolic and not Dirac-like [59].

Contrary to the VPS in a smooth THJ, these other massive interface states (ECS and DKS) appear only at an abrupt THJ and do not exist if the band bending is lifted or the bias removed. In particular, an applied electric field cannot induce a gradual $\Delta(z)$ band inversion as required for the VPS and for these reasons the VPS do not need any external perturbation to appear in a gradual THJ. It is noted that the strong electrostatic potential required to observe ECS or DKS intrinsically induces a strong Rashba splitting that is not present for the VPS and this Rashba splitting drastically affects the ECS and DKS dispersions. Moreover, the response of the ECS and DKS states towards an applied electrostatic field is different [57]: whereas VPS are only shifted upward or downward in energy (depending on the sign of the field), and even *disappear* above a certain critical field [50] (see Figs. 8 and 9 of Appendix D), on the contrary, ECS and DKS *emerge* when the electric field increases in a definite direction due to the enhanced of quantum confinement with increasing band bending [56]. The nature of the VPS confinement is therefore fundamentally different from the ECS and/or DKS states. This also follows from the fact that while the ESC and DKS are essentially confined to only one side of the topological heterojunction, i.e., their wavefunctions are localized in a region of constant band topology, whereas those of the VPS nearly equally extend over both TI and NI regions as shown by Fig. 3.

Last but not least, the ECS and DKS do not display the characteristic electron-hole symmetry of the VPS, as for an upward or downward band bending the ESC states are formed only either in the valence or in the conduction band; and the parabolic DKS are highly asymmetric due to the mixing with the heavy hole band, and form only electron-like states at the HgTe/HgCdTe interface for instance [59]. Indeed, only THJs made of 3D Dirac materials such as PbSn(Se,Te)are described by a Hamiltonian in a supersymmetric form [8] which leads to a pure relativistic spectrum of electron-hole symmetric Weyl and Dirac states as demonstrated in this work. For all these reasons, the ESC and DKS are not to be confused with the VPS that are of fundamentally different origin.

## V. CONCLUSION

In conclusion, we have shown that the smooth interconnection of materials with different band topologies exposes a rich relativistic spectrum of coexisting 2D chiral Weyl and massive Dirac Volkov-Pankratov states that are both confined to the same interface. The Volkov-Pankratov states emerge only for a sufficiently smooth topological heterojunctions and they expose an energy spectrum that is controlled on-demand by the interface width. Using magneto-spectroscopy of engineered normal insulator/topological crystalline insulator heterojunctions based on the perfect Dirac $PbSe/Pb_{1-x}Sn_xSe$ system, we demonstrated the tunability of the energy spectrum with Dirac VP states quantized in terms of $\sqrt{n/l}$, as well as their coexistence with the zero energy massless Weyl fermions that persist for arbitrary interfaces. Our work thus offers an unequivocally experimental validation of the original Volkov-Pankratov theory. We also highlight the different nature of the gapless chiral state at a gradual topological interface where its Gaussian envelope function strongly differs from that of the cusped shape at abrupt interfaces. The demonstrated basic principle of topological band profile engineering of THJ can be generalized and applied to other topological systems such as $(Bi_{1-x}In_x)_2Se_3$ or $BiTl(S_{1-\delta}Se_\delta)_2$ where similar smooth topological transitions can be realized by grading of composition [62,63]. Engineering of the topological interface profiles thus, opens the doors for further exploration of other non-linear gap variations that could simulate different relativistic particles and/or effects. By establishing a textbook platform to access Weyl and Dirac fermions, our work paves the way for the investigation of various symmetry breaking effects on multiple relativistic particles by introducing magnetism or ferroelectric polarization that can be induced in the IV-VI TCIs, e.g., by Mn or Ge doping [1,33,64,65]. Last but not least, VPS have been theoretically predicted to be encountered in various other scenarios such as in graphene nanoribbons [66], in partial dislocations in bilayer graphene [67], as well as in topological superconductor – superconductor heterojunctions, so that VPS can emerge alongside with Majorana fermions [68].

## ACKNOWLEDGMENTS

We thank M. Goerbig and R. Ferreira for fruitful discussions, as well as L. Petaccia and N. Olszowska for their supports at the ELETTRA and SOLARIS synchrotrons, respectively. The authors acknowledge the funding support of the ANR (Contract N° ANR-19-CE30-022-01) and the Austrian Science Funds

FWF (Project I-4493). The work at SOLARIS was developed under the provision of the Polish Ministry of Education and Science project: "*Support for research and development with the use of research infrastructure of the National Synchrotron Radiation Centre SOLARIS*" under contract nr 1/SOL/2021/2. We acknowledge SOLARIS Centre for the access to the Beamline "UARPES", where the measurements were performed.

## APPENDIX A: TOPOLOGICAL HETEROJUNCTION GROWTH

The IV-VI based THJ were grown by MBE on (001) KCl substrates. Layers with incrementally varying Pb$_{1-x}$Sn$_x$Se compositions were grown on top of each other using PbSe and SnSe effusion sources. The Sn content in each layer was controlled by adjusting the SnSe:PbSe beam flux ratio that was measured accurately by the quartz microbalance method. The carrier concentration in the THJs was controlled using a Bi$_2$Se$_3$ source to induce a slightly n-doping of the Pb$_{1-x}$Sn$_x$Se layers in the THJ and compensate the native *p*-type character of the material that increases with increasing Sn content. Growth was carried out at substrate temperatures around 300°C and ultra-high vacuum conditions (<5x10$^{-10}$ mbar). 2D growth was sustained throughout the whole growth process as checked by in situ high-energy electron diffraction.

To deliberately form a slowly varying Sn gradient across the THJs, the graded junction region was split up in 9 successive subregions each consisting of an ultra-short period superlattice (SL) consisting of PbSe and Pb$_{0.76}$Sn$_{0.24}$Se layers with varying thickness ratio as shown schematically in the Fig. 5. The period of these SLs was fixed to ~2.5 nm and the total thickness of each SL stack was ~10 % of the designated entire junction width. The increment in the average Sn content in successive SL was therefore 2.4 %, but these incremental steps were further smoothed by Sn/Pb interdiffusion. For the thicker THJ-150 and THJ-200, the steps in the Sn increments was reduced to 2 and 1.5 %, respectively.

The NI PbSe and TCI Pb$_{0.76}$Sn$_{0.24}$Se reference layers used for ARPES experiments were grown under the same conditions and consist of 400 nm thick epilayers. However, they were higher *n*-doped to ensure a Fermi energy sufficiently high in the conduction band in order to observe the conduction band in the ARPES measurements.

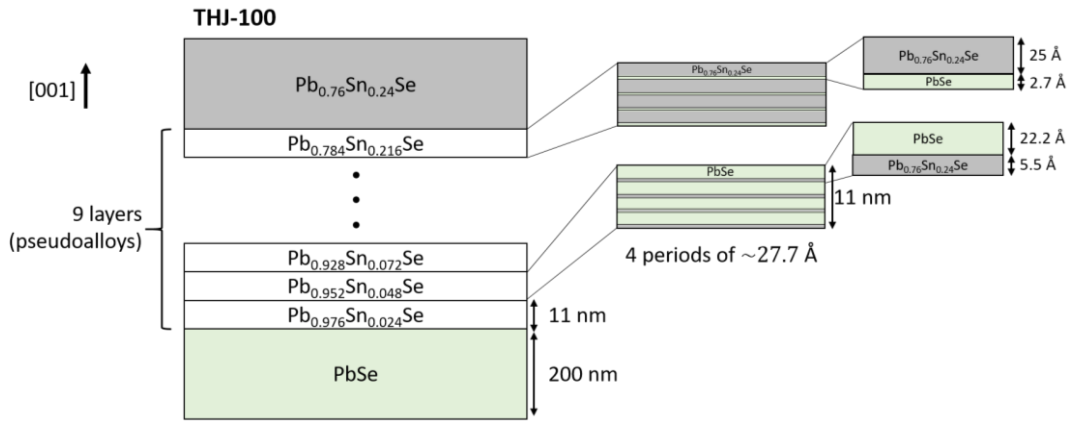

**FIG. 5.** Scheme of the THJ growth, with the deliberate formation of the gradual interface. The values shown here correspond to THJ-100. Similar values were used for the other THJs.

## APPENDIX B: MAGNETO-OPTICAL CHARACTERIZATION OF THE THJs

Magneto-optical measurements in Faraday geometry were performed on the six investigated THJ samples of this work. Figure 6 shows the magneto-optical results obtained on sample THJ-100, which stand for a typical result for the six investigated THJs in the main text. The transmission spectra plotted in Fig. 6(a) show several absorption lines that shift to higher energies when the magnetic field is increased, revealing optical transitions between Landau levels. The absorption energies, marked by colored arrows, are extracted and their variation versus magnetic field are represented by the colored dots in Fig. 6(b). A fit of the high energy part (>100 meV) is achieved by using the expression of massive Dirac Landau levels corresponding to those of 3D $Pb_{1-x}Sn_xSe$ alloys [14,40,69]:

$$E_{N,\pm,\xi} = \frac{2\xi M_{TCI/NI} eB}{\hbar} \pm \sqrt{\left(\Delta_{TCI/NI} + \frac{2M_{TCI/NI} eBN}{\hbar}\right)^2 + 2e\hbar v_{\parallel,TCI/NI}^2 BN} \quad (5)$$

where $N \geq 0$ is the Landau level index, $\xi = \pm 1$ denotes the two spin orientations. The inversion parameter $M_{TCI/NI}$ accounts for the effect of all other remote bands, which is rather small as far-bands are typically 2 eV above and below the Dirac point [15]. $\Delta_{TCI/NI}$, $v_{\parallel,TCI/NI}$ and $M_{TCI/NI}$ are fitting parameters which depend on the chemical composition and temperature of the $Pb_{1-x}Sn_xSe$ alloy [14,69].

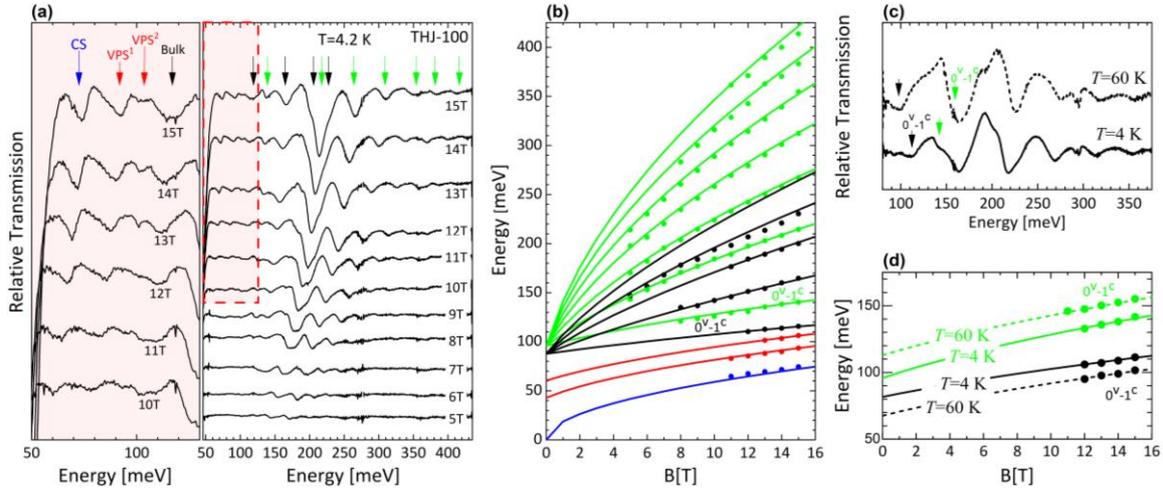

**FIG. 6.** **(a)** Normalized magneto-optical transmission spectra of THJ-100 at $T = 4.2$ K and for different magnetic fields. The colored arrows indicate the observed absorptions at $B = 15$ T. Left: zoom of the low energy part at high magnetic field, corresponding to the area shown inside the dashed red line on the right panel. **(b)** Full magneto-optical fan chart of THJ-100, where the absorption energies versus magnetic field are represented by dots. The solid lines are the calculated transitions between Landau levels using Eq. (5). In green and black the absorptions and transitions are attributed to PbSe and $Pb_{0.76}Sn_{0.24}Se$, respectively. **(c)** Spectra at $B = 15$ T at $T = 4.2$ K (solid line) and $T = 60$ K (dashed line). **(d)** Fit of the ground transitions $0^v$-$1^c$ (or $|E_{1,+} - E_{0,-}|$ following Eq. (5)) attributed to the two thick epilayers at $T = 4.2$ K (solid lines) and $T = 60$ K (dashed lines).

The solid lines in Fig. 6(b) represent the calculated transition energies, using the parameters given in Table I, occurring between Landau levels in Faraday geometry for $B//[001]$: $|E_{N\pm1,+} - E_{N,-}|$ at fixed $\xi$. The fits of the experimental data (green and black dots) involve two Landau level spectra, both described by Eq. (5). They correspond to the two thick epilayers that sandwich the gradual interface: PbSe and $Pb_{0.76}Sn_{0.24}Se$. The fits lead to the three band parameters for each layer: $2\Delta_{TCI/NI}$ is extracted by the experimental extrapolation of the absorptions at $B = 0$; $v_{\parallel,TCI/NI}$ and $M_{TCI/NI}$ account for the slope of the transitions versus $B$. The band parameters giving the best fits are listed in Table I. The black series presents the smallest velocity value $v_{\parallel,TCI}$ (slowly dispersive transitions) and is thus associated to $Pb_{0.76}Sn_{0.24}Se$, whose high Sn content is responsible for its topological character[23,32]. In this sample, the trivial PbSe layer (green series) has a gap of $+95$ meV and the one of the topological $Pb_{0.76}Sn_{0.24}Se$ layer is $2\Delta_{TCI} = -90$ meV.

To confirm the trivial and topological character of each layer, temperature dependent measurements on the same THJ sample are performed. The spectra at $B = 15$ T from $T = 4.2$ K to $T = 60$ K (see Fig.

6(c)) show a shift of the absorptions toward high energies for the green series and towards low energies for the black series. As it is shown by the corresponding fits presented in Fig. 6(d), this effect is due to an increase of the absolute gap value for PbSe while it is decreasing for $Pb_{0.76}Sn_{0.24}Se$ (see the extrapolations at $B = 0$). This temperature dependence of the absolute energy gap is a direct evidence of the inverted band structure of $Pb_{0.76}Sn_{0.24}Se$ [21,70]. Therefore, this system realizes a heterojunction with two topologically different epilayers having gaps of nearly similar absolute value but with opposite sign. This data corroborates and supports the ARPES measurements exposed in the main text.

**Table I.** Energy gap and in-plane Dirac velocities of the trivial and topological thick epilayers that sandwich the gradual interface. The magneto-optical determination yields error bars of $\pm 5$ meV on the gaps and $\pm 0.05 \times 10^5$ m/s for the velocities. The inversion parameters are fixed to $M_{NI} = 13$ eV.Å$^2$ and $M_{TCI} = 19$ eV.Å$^2$.

| Samples | $2\Delta_{NI}$ [meV] | $2\Delta_{TCI}$ [meV] | $v_{\parallel,NI}$ [x10$^5$ m/s] | $v_{\parallel,TCI}$ [x10$^5$ m/s] |
|---|---|---|---|---|
| THJ-0 | 90 | -90 | 5.3 | 4.4 |
| THJ-50 | 95 | -90 | 5.4 | 4.4 |
| THJ-75 | 90 | -90 | 5.3 | 4.4 |
| THJ-100 | 95 | -90 | 5.3 | 4.4 |
| THJ-150 | 110 | -100 | 5.2 | 4.4 |
| THJ-200 | 100 | -85 | 5.3 | 4.4 |

## APPENDIX C: $k.p$ MODELLING OF A THJ

Beyond the analytical formula Eq. (3), numerical $k.p$ calculations in the envelope function approximation have been performed to unravel the relativistic spectrum of a THJ. Using Hamiltonians developed in Refs. [40–42,71], and substituting $k_z \rightarrow -i\hbar \frac{\partial}{\partial z}$ to account for the confinement, we have:

$$\begin{pmatrix} -\Delta(z) - E & \xi i\hbar v_z \frac{\partial}{\partial z} \\ \xi i\hbar v_z \frac{\partial}{\partial z} & \Delta(z) - E \end{pmatrix} \begin{pmatrix} F_1^{(j)} \\ \xi F_2^{(j)} \end{pmatrix} = 0 \qquad (6)$$

where $\xi = \pm 1$ denotes again the Kramers pairs. The system (6) evolves in:

$$\begin{cases} -(\Delta(z) + E_j)F_1^{(j)} + \hbar^2 v_z^2 \frac{d}{dz} \frac{1}{\Delta(z) - E_j} \frac{dF_1^{(j)}}{dz} = 0 \\ F_2^{(j)} = -\frac{i\hbar v_z}{(\Delta(z) - E_j)} \frac{dF_1^{(j)}}{dz} \end{cases}$$

Let us solve this problem for an abrupt $\Delta(z)$ potential. If it is so, one can write $F_1^{(j)}$ as a sum of an in-coming and out-coming wave on both sides of the abrupt $\Delta(z)$ variation. These waves can be evanescent or propagative depending on the sign of $\Delta(z) - E_j$. Then, at the abrupt interface, the solutions found on both sides are matched so that $F_1^{(j)}$ and $\frac{1}{\Delta(z)-E_j} \frac{dF_1^{(j)}}{dz}$ are continuous.

To model a THJ, the smooth interface can be divided into $N + 1$ parts where $\Delta_p$ is independent from $z$ with $0 \leq p \leq N$ and $N$ the number of abrupt interfaces. The smooth gradual potential is thus modelled by a succession of $N$ abrupt interfaces separating regions of constant gaps $\Delta_p$. When $N$ tends to infinity, we retrieve a smooth potential. Usually, $N = 100$ insures a good convergence of the numerical computation. On one step of the staircase profile, the solutions are known (see above) and one can write the spinor wavefunction as a sum of in-coming and out-coming waves:

$$\varphi_p(z) = A_p \begin{pmatrix} \frac{\hbar v_z k_p}{\Delta_p + E} \\ 1 \end{pmatrix} e^{ik_p z} + B_p \begin{pmatrix} -\frac{\hbar v_z k_p}{\Delta_p + E} \\ 1 \end{pmatrix} e^{-ik_p z}$$

with the wavevector:

$$k_p = \begin{cases} \frac{1}{\hbar v_z}\sqrt{E^2 - \Delta_p^2}, & |E| > |\Delta_p| \\ \frac{i}{\hbar v_z}\sqrt{\Delta_p^2 - E^2}, & |E| < |\Delta_p| \end{cases}$$

$A_p$ and $B_p$ are constants that are determined by using the continuity conditions at the $N$ interfaces. When the wavefunction is imposed to vanish far from the global gradual interface, i.e., in the PbSe and $Pb_{0.76}Sn_{0.24}Se$ thick epilayers, only a few solutions remain at energies corresponding to VPS and CS. The wavefunction is then normalized.

The major advantage of this resolution is that the staircase function $\Delta_p$ is obtained by taking the targeted $\Delta(z)$ potential and substituting $z$ by $l\left(\frac{2p}{N} - 1\right)$. It is therefore easy to consider other non-linear potential profiles. Fig. 7 shows examples of different relativistic VP energy spectra, obtained with three distinct potential profiles: linear, hyperbolic tangent and exponential.

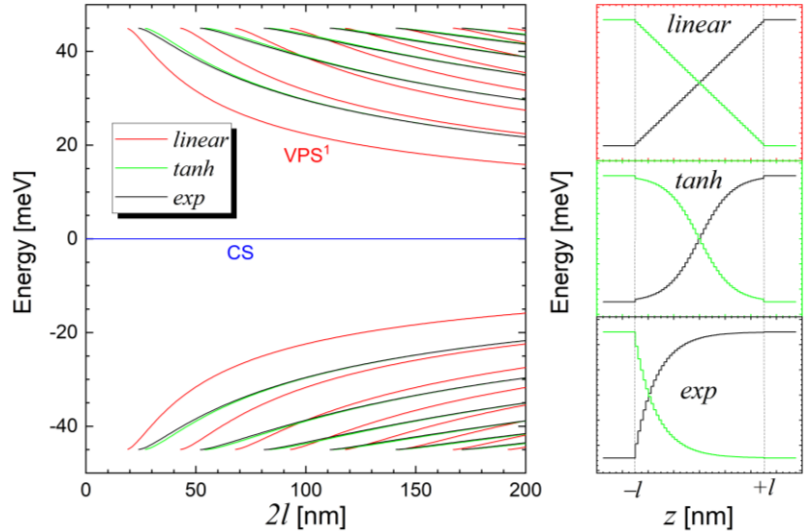

**FIG. 7.** Calculated CS and VPS energies versus interface thickness $2l$, with the parameters listed in Table I and with different potential profile: linear (red), hyperbolic tangent (green) and exponential (black).

### APPENDIX D: THJ UNDER ELECTROSTATIC POTENTIAL

One can take into account the effect of an electrostatic potential on a THJ by adding a term $eFz\mathbb{I}$ in Eq. (6), which can be due to a band misalignment or an applied electric field for instance. The $L_6^+$ and $L_6^-$ band edges are then shift together linearly over the thickness $2l$. The calculations are performed following the method given in Appendix C, by segmenting $eFz$ into $\frac{2eFlp}{N}$. THJ-100 energy spectra are given in Fig. 8 for five different $F$ values. The corresponding CS and VPS square wavefunctions are also represented.

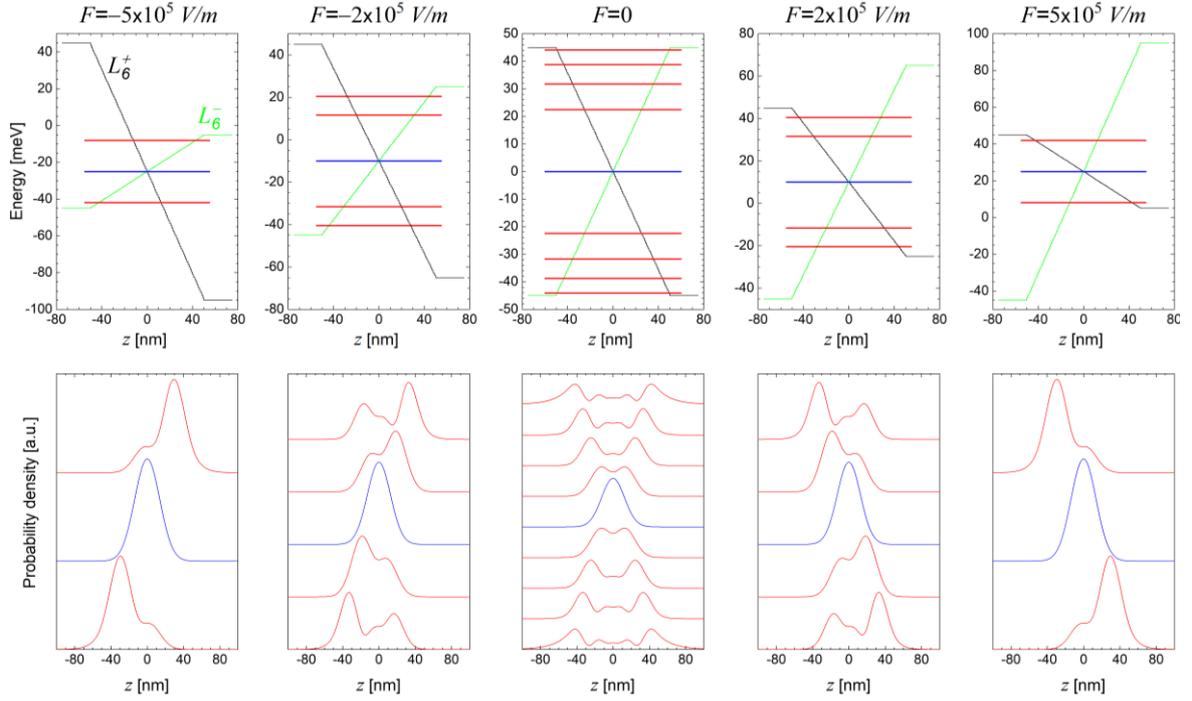

**Fig. 8.** THJ-100 calculated spectra along $z$ for five different values of the electric field $F$. The $L_6^+$ and $L_6^-$ band edges are drawn in black and green. The energy of the confined states at the interface are indicated in blue for the CS and in red for the VPS. Their corresponding square wavefunctions are calculated and shown below.

The calculations presented in Fig. 8 show the remarkable conservation of the electron-hole symmetry of the VPS, which is specific for a relativistic spectrum. They also show the breakdown of the VPS with electric field. These remarks are also illustrated by Fig. 9, where the energy evolution of the CS and VPS versus $F$ are represented. Similar effects have been theoretically investigated and more detailed in Ref. [50]. Note that the VPS gaps remain steady when the electric field is small, meaning that a slight band misalignment or a weak charge transfer would not have any influences on our magneto-optical VPS gaps determination.

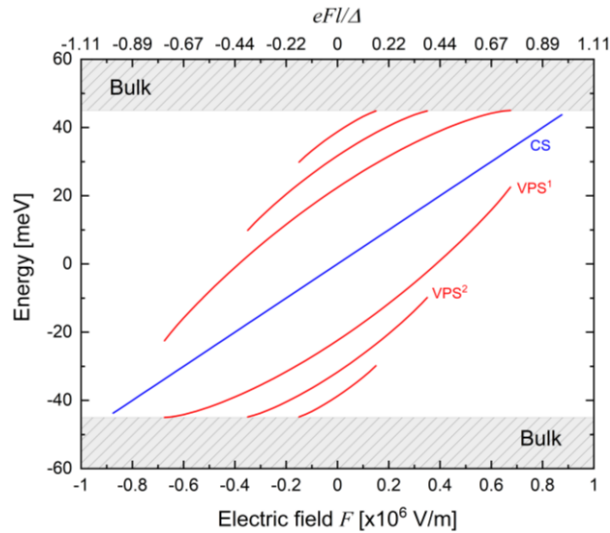

**Fig. 9.** Calculated energy evolution of the CS (in blue) and the VPS (in red) versus the electric field $F$ in THJ-100.

## APPENDIX E: THJ MAGNETO-SPECTROSCOPY WITH IN-PLANE MAGNETIC FIELD

We performed a magneto-optical experiment where the sample THJ-75 was placed parallel to the magnetic field direction, i.e, $\boldsymbol{B}$ is in the (001) plane. The spectra at $B = 15$ T is shown in Fig. 10. Whereas the high energy part ($2\Delta_{TCI,NI} \gtrsim 100$ meV), which hosts the absorptions minima due to 200 nm thick epilayers PbSe and $Pb_{0.76}Sn_{0.24}Se$, still exhibits inter Landau level absorptions; the low energy part (highlighted in red) shows no absorption minima. This behavior is expected for 2D states when the magnetic field is applied perpendicularly to the confinement direction. Indeed, no Landau quantization occurs in this geometry, i.e., the dispersions remain conical, and thus, no transitions between Landau levels appear. Moreover, the density of states is not increased for higher magnetic fields as it is the case with a Landau quantization, where each level displays a $eB/h$ degeneracy. Therefore, the Fermi level remains above the CS and VPS as they are located within the gap of the THJ two ends PbSe and $Pb_{0.76}Sn_{0.24}Se$. As a result, no transitions between states located at the band edges are allowed due to a population effect. If the Fermi level of the investigated THJs was lower, the absorptions between VPS or CS would have the shape of their densities of state, thus, we would expect a step (for VPS) and/or a line (for CS) in the transmission spectra [49]. This is not observed here.

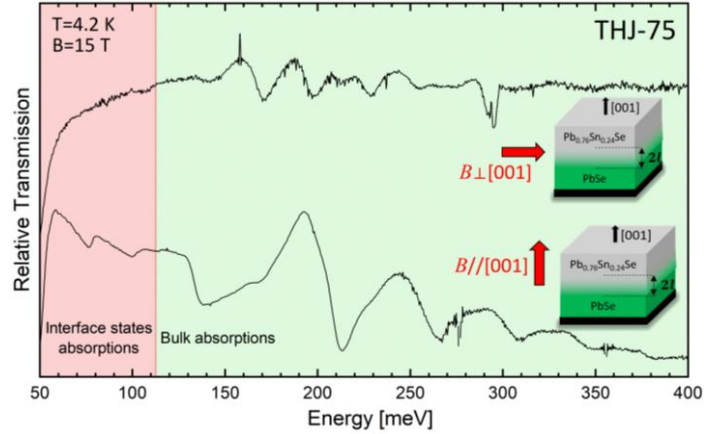

**FIG. 10.** Spectra of THJ-75 at $T = 4.2$ K and $B = 15$ T for two different geometries. Upper spectrum: $\boldsymbol{B}$ is in the plane of the sample, lower spectrum: Faraday geometry ($\boldsymbol{B}$ perpendicular to the sample, i.e., parallel to [001]). The high and low energy parts are shaded in distinct colors.